\documentclass[epsf]{article}

\usepackage{graphicx}
\usepackage{amssymb}
\usepackage{amsmath}
\usepackage{latexsym}

\def\abstract#1{\vskip 7mm 
        \begin{center}{\large Abstract}\par \smallskip
                \begin{minipage}[c]{12cm}
                        \small #1
                \end{minipage}
        \end{center}
}
\def\title#1{\begin{center}{\Large\bf #1}\end{center}}
\def\author#1{\vskip 5mm \begin{center}{#1}\end{center}}
\def\address#1{\begin{center}{\it #1}\end{center}}
\makeatletter
\@ifundefined{lesssim}{}{}
\@ifundefined{gtrsim}{}{}
\def\vereq#1#2{\lower3pt\vbox{\baselineskip1.5pt \lineskip1.5pt
\ialign{$\m@th#1\hfill##\hfil$\crcr#2\crcr\sim\crcr}}}
\makeatother

\begin{document}

\title{%
   Non-linear Gravity on Branes and Effective Action
}
\author{%
  Jiro Soda,\footnote{E-mail:jiro@phys.h.kyoto-u.ac.jp}
  Sugumi Kanno \footnote{E-mail:kanno@phys.h.kyoto-u.ac.jp}
}

\address{ 
  ${}^1$Department of Fundamental Sciences, FIHS, 
Kyoto University, Kyoto 606-8501, Japan
}
\address{%
${}^2$
Graduate School of Human and Environmental Studies, Kyoto University, Kyoto
 606-8501, Japan 
 }
\abstract{
  We  develop the general formalism to study the low energy regime of the
 brane world. We apply our formalism to the single brane model
 where the AdS/CFT correspondence will take an important role.
 We also consider the two-brane system and show the system
 is described by the quasi-scalar tensor gravity. 
 Our result provides a basis for predicting CMB fluctuations 
 in the braneworld models. 
}

\section{Introduction}

The existence of the initial singularity in the standard cosmology is a 
 notorious problem which is expected to be solved by taking into account 
 the quantum effects of the gravity. 
It is widely accepted that the superstring theory is the most promising
 candidate for the quantum theory of gravity. One prominent feature of the
 superstring theory is the existence of the extra dimensions. 
 To fill the gap between this theoretical prediction and our 4-dimensional
 universe, we need to hide the extra dimensions someway. 
 The conventional idea is the Kaluza-Klein compactification scenario 
 where the internal dimensions are assumed to be compactified to the Planck 
 scale.
 Recently, however, a new picture, the so-called braneworld, has emerged
 thanks to the developments of the non-perturbative aspects of the superstring
 theory~\cite{braneworld,RS1}. 
 In the braneworld picture, the ordinary matter exists on the
 brane while the gravity can propagate in the bulk. In particular, 
 we are on the brane! Hence,
 what we would like to know is how the non-linear gravity appears
 on the brane. In general, it would be difficult to get such a description
 due to the strong coupling of the bulk degrees of freedom with those of
 the brane. However, in the low energy regime ${\rho \over \sigma} 
    \sim \ell^2 R \ll 1$, it is possible to obtain the 4-dimensional 
effective theory approximately~\cite{wiseman,kanno1,kanno2}.
 Indeed, this is sufficient except for the
 extreme situation for which we need more profound understanding 
 of the string theory.  For example, if we take the curvature scale
 $\ell$ as $10^{-16}{\rm cm}$, our approximation is valid at
  the energy scale below $ 10^{11}{\rm GeV} $ or  the gravitational 
  radius greater than $ 10^{-21}{\rm km} $.  
  In fact, this is the interesting regime for most
 astrophysical and cosmological phenomena. 
 In this paper, we review our recent results on this 
 issue~\cite{kanno1,kanno2}. 
 
 In the next section, we will develop the general
 formalism. In sec.3, we apply our formalism to the single brane model
 where the AdS/CFT correspondence will play an important role.
 In sec.4, we consider the two-brane system and show the system
 is described by the quasi-scalar tensor gravity. 
 Sec.5 is devoted to the conclusion.

\section{General Formalism}  

 If there exists no matter, the ground state is the Minkowski spacetime
  in the 4-dimensional theory. 
 Correspondingly, if there exists no matter on the  brane, we would expect
the induced metric on the brane is Minkowski and the bulk 
 geometry is   the Anti-deSitter spacetime. Indeed, we have
 such  solution 
\begin{eqnarray*}
ds^2 = dy^2 + \Omega^2 (y) \eta_{\mu\nu} dx^{\mu} dx^{\nu} \ ,
\end{eqnarray*}
where $ \Omega^2 = \exp[ -2 {\displaystyle{ y/\ell}} ]$ is the 
warp factor. The brane is located at $y=0$ in this coordinate system. 
 To obtain the above solution we have imposed the relation $\kappa \sigma
 = 6/\ell$. 
 
Let us put the small amount of  matter on the brane. Then, the brane 
will be curved and the bulk geometry will be deformed as
\begin{eqnarray}
ds^2 = dy^2 
	+ \left( \Omega^2 (y) h_{\mu\nu} (x) 
	+ \delta g_{\mu\nu} (y,x^{\mu} ) \right) 
	dx^{\mu} dx^{\nu} \ ,
\end{eqnarray}
where the boundary condition $\delta g_{\mu\nu} (y=0 ,x^{\mu} ) =0 $
 is imposed so that $h_{\mu\nu}$ becomes the induced metric on the brane.
How the geometry will be deformed is determined by the 5-dimensional
 Einstein equations: 
\begin{eqnarray}
 G^{(5)}_{AB} = { 6\over \ell^2} g_{AB} 
    + \delta (y) 8\pi G_N \ell
    \left( -\sigma g_{\mu\nu} +T_{\mu\nu} \right) \delta^\mu_A \delta^\nu_B   
    \ , \quad A = (y,\mu) \ ,
\end{eqnarray}
where $T_{\mu\nu}$ is the energy-momentum tensor of the matter.
As we are considering the deviation from the Anti-deSitter spacetime,
 it is convenient to define the variables
\begin{eqnarray}
 \delta K^{\mu}_{\ \nu} = - {1\over 2} \delta \left[ 
                           g^{\mu\alpha} g_{\alpha\nu ,y} \right]
\equiv \delta \Sigma^\mu_{\ \nu} + {1\over 4} \delta^\mu_{\ \nu} 
  \delta K  \ ,
\end{eqnarray}
where $\delta \Sigma^\mu_{\ \mu} =0$.
 In terms of these variables, the Hamiltonian constraint equation
 becomes
\begin{eqnarray}    
 \delta K  = - {\ell \over 6} \left[
    {3\over 4} \delta K^2 - \delta \Sigma^\mu_{\ \nu}
    \delta \Sigma^\nu_{\ \mu}  - \stackrel{(4)}{R} \right] 
\end{eqnarray}
and the momentum constraint equation reads 
\begin{eqnarray}    
 \nabla_\lambda  \delta \Sigma^\lambda_{\ \mu}
      - {3\over 4} \nabla_\mu \delta K =0  \ .   
\end{eqnarray}
The evolution equation in the direction to $y$ is
\begin{eqnarray}
 {1\over \Omega^4} \left[ \Omega^4 \delta \Sigma^\mu_{\ \nu} \right]_{,y} 
    = \delta K \delta  \Sigma^\mu_{\ \nu}
    - \left[ {\stackrel{(4)}{R^\mu_{\ \nu}}} \right]_{\rm traceless } 
    \ .
\end{eqnarray}
As we have the singular source at the brane position, 
we must take into account the junction condition, 
\begin{eqnarray}
  \displaystyle{2\over \ell} \left[ \delta \Sigma^\mu_{\ \nu} - 
 {3\over 4} \delta^\mu_\nu \delta K \right] \Bigl|_{y=0} 
    = 8\pi G_N  T^\mu_{\ \nu}  \ ,
\end{eqnarray}
where we have imposed the $Z_2 $ symmetry on the spacetime. 

After solving the bulk equations of motion, the junction condition
 gives the effective equations of motion for the induced metric.
 By integrating the evolution equation, we have 
\begin{eqnarray*}
  {2\over \ell}\left[\delta \Sigma^\mu_{\ \nu}
  -{3\over 4} \delta^\mu_\nu \delta K \right] 
  &=&  -{\ell^2 \chi^\mu_{\ \nu} \over \Omega^4}
  -{2\over \ell \Omega^4 }\int^y dy 
  \Omega^4 \left[ {\stackrel{(4)}{R^\mu_{\ \nu}}} 
  - {1\over 4} \delta^\mu_\nu \stackrel{(4)}{R} 
  -\delta K \delta \Sigma^\mu_{\ \nu} \right]
  \\
 &&  - {1\over 4} \delta^\mu_\nu \stackrel{(4)}{R} 
  +{1\over 4}\delta^\mu_\nu 
  \left[ {3\over 4} \delta K^2 - \delta \Sigma^\mu_{\ \nu}
    \delta \Sigma^\nu_{\ \mu} \right] \\
  &=&  -{\ell^2 \chi^\mu_{\ \nu} (x) \over \Omega^4 (y)} 
    + \stackrel{(4)}{G^\mu_{\ \nu}} (\Omega^2 (y) h_{\mu\nu} (x) ) 
     + {\cal O} (\ell^4 R^2 ) \ ,
\end{eqnarray*}
where we introduced the constants of integration $\chi_{\mu\nu}$.
 This integration can be performed iteratively with the expansion
 parameter $\ell^2 R$. 
The resulting equations take the form
\begin{eqnarray}
     \stackrel{(4)}{G^\mu_{\ \nu}} (
     \Omega^2 |_{y=0} h_{\mu\nu} )
     = 8\pi G_N T^\mu_{\ \nu}  +  
      {\ell^2 \over \Omega^4 |_{y=0}} 
         \chi^\mu_{\ \nu}   +  t^\mu_{\ \nu }  \ ,
\end{eqnarray}
where we have included the trivial factor $ \Omega |_{y=0} =1 $ 
 for memorizing how the warp factor comes in to the effective theory.
 Here we have decomposed the corrections to the conventional Einstein
 theory into the nonlocal part $\chi_{\mu\nu}$ 
 and the local part $t_{\mu\nu}$.

We can  expand these corrections in the order of the $\ell^2 R$:
\begin{eqnarray}
  {\rm Nonlocal} \quad
   \chi_{\mu\nu} &=& \underbrace{\chi^{(1)}_{\mu\nu}}_{
       \mbox{$ {\cal O} (\ell^2 R) $}} +\underbrace{\chi^{(2)}_{\mu\nu}}_{
       \mbox{$ {\cal O} (\ell^4 R^2 ) $}} + \cdots \\
 {\rm Local} \quad
        t_{\mu\nu} &=& \qquad \qquad \underbrace{t^{(2)}_{\mu\nu}}_{
       \mbox{$ {\cal O} (\ell^4 R^2 ) $}} + \underbrace{t^{(3)}_{\mu\nu}}_{
       \mbox{$ {\cal O} (\ell^6 R^3 ) $}}  + \cdots    
\end{eqnarray}
where the expansion of the local tensor starts from the second order
 because the first order part is already included as the Einstein tensor. 
 Notice that we have
$
\chi^{(1)\mu}_{\quad \  \mu} = 0 
$ 
because of  $\Sigma^\mu_{\ \mu }=0$.

\section{Single Brane Model (RS2): AdS/CFT correspondence}

Now we shall apply the general formalism to the single brane 
model~\cite{kanno1}.
 A natural boundary condition for the single brane model is to
 impose the regularity at the Cauchy horizon, namely asymptotically
 AdS boundary condition.  Taking this boundary condition, 
 we obtain $\chi^{(1)}_{\mu\nu} =0 $. Thus, we have recovered
 Einstein theory at the leading order.

 At the next order ${\cal O} ( l^4 R^2 ) $, 
 the traceless part of $t^{(2)}_{\mu\nu}$ is proportional to
\begin{eqnarray}
   {\cal S^\mu_\nu} 
     &=& R^\mu_{\ \alpha} R^\alpha_{\ \nu}
             -{1\over 3} R R^\mu_{\ \nu} 
         -{1\over 4} \delta^\mu_\nu (R^\alpha_{\ \beta} R^\beta_{\ \alpha}
         - {1\over 3} R^2)  \nonumber \\ 
    & & \quad     -{1\over 2} \left( R^{\alpha\mu}_{\ \ |\nu\alpha}
                    + R^{\alpha \ |\mu}_{\ \nu \ \  | \alpha}  
              -{2\over 3} R^{|\mu}_{\ |\nu}  - \Box R^\mu_{\ \nu} 
              +{1\over 6} \delta^\mu_\nu \Box R \right)   
\end{eqnarray}
where ${\cal S}^{\mu}_{ \ \nu | \mu}$ is transverse and traceless: 
$
  {\cal S}^{\mu}_{ \ \nu | \mu} =0  \ ,  
  \  {\cal S}^{\mu}_{ \ \mu} = 0  \ .  
$
 The nonlocal part $\tilde{\chi}^{(2)}_{\mu\nu} 
 = \chi^{(2)}_{\mu\nu} + 1/4 h_{\mu\nu} t^{(2)\mu}{}_{\mu}$ 
 can not be determined by the Einstein equations, but  constrained as 
$$
  \tilde{\chi}^{(2)\mu}_{\quad \mu}  
     =-{1\over 8} \left( R^\alpha_{\ \beta} R^\beta_{\ \alpha} 
            - {1\over 3}  R^2 \right) \ .      
$$
Here, the AdS/CFT correspondence comes in. As the trace anomaly for some
 supersymmetric theories proportional to the above result, 
 we make the following identification
\begin{eqnarray}
\tilde{\chi}^{(2)}_{\mu\nu} 
      = { \kappa^2\over \ell^3} T_{\mu\nu}^{\rm CFT} \ . 
\end{eqnarray}
 Note that the  effective number of the CFT fields  
$
{\ell^3 / \kappa^2} \sim {\ell^2 / G} \sim 10^{66} 
$
is so huge. This is the regime where AdS/CFT correspondence holds.

Thus, the effective equations of motion become
\begin{eqnarray}
     G^{(4)}_{\mu\nu} = 8\pi G_N T_{\mu\nu} + 8\pi G_N T_{\mu\nu}^{\rm CFT}
         + \alpha  {\cal S}^{\mu}_{\ \nu } \ . 
\end{eqnarray}
Seeing Eq.(11), one can read off the effective action  
\begin{eqnarray}
   S_{\rm eff}  = {1\over 16\pi G_N } \int d^4 x \sqrt{-h} R
                    + S_{\rm matter} + S_{CFT}   
       +  { \alpha \ell^2 \over 16 \pi G_N}
     \int d^4 x \sqrt{-h}
           \left[R^{\mu\nu} R_{\mu\nu}
                   - {1\over 3}  R^2 \right]   \ . 
\end{eqnarray}

 Now one can consider the cosmology using the above effective theory.
One can obtain the renormalized action for the CFT, $S^{\rm CFT}$, then
 we can deduce the one point function from the formula
$$
     <T^{\rm CFT}_{\mu\nu}> = -{2\over \sqrt{-g}} 
     {\delta S^{\rm CFT} \over \delta g^{\mu\nu}}  \ .
$$
The two point correlation function can be also calculated as 
$$
   <T^{\rm CFT}_{\mu\nu} (x) T^{\rm CFT}_{\lambda\rho}(y) > 
       = -{2\over \sqrt{-g}} {\delta <T^{\rm CFT}_{\mu\nu} (x) >
        \over \delta g^{\lambda\rho} (y) }   \ .
$$
Thus, we obtain the  perturbed effective Einstein  Equations
\begin{eqnarray}
  \delta G_{\mu\nu} = 8\pi G_N \delta T_{\mu\nu} 
       -{1\over 2} \int d^4 y \sqrt{-g(y) } 
       <T^{\rm CFT}_{\mu\nu} (x) T^{{\rm CFT}\lambda\rho}(y) >
        \delta g_{\lambda\rho}
       + \alpha \delta t^{(2)}_{\mu\nu} \ .
\end{eqnarray}
This is nothing but the integro-differential equation. 
 It is possible in principle to solve numerically the linearized
 equations of motion in the cosmological situation. 
 We leave this for the future work.

\section{Two Brane Model (RS1): Radion }

In this section, we will consider the two-brane system
 which is more realistic from the M-theory point of 
 view~\cite{wiseman,kanno2}(see also \cite{kanno3}). 
In the two-brane system, the radion field plays an important role. 
The radion is defined as the distance between the positive tension
 brane and the negative tension brane, $ d(x) $. 
Now the  warp  factor $ \Omega^2 = \exp[-2{d(x)/ \ell}] $
 becomes dynamical variable. 
 
 The general formula gives the equation on the positive tension 
brane:  
\begin{eqnarray}
     G^{(4)\mu}_{\quad \ \nu} (h_{\mu\nu} ) 
     = {\kappa^2 \over \ell} T^{\oplus \mu}_{\quad \ \nu} 
                          + \ell^2 \chi^{(1)\mu}_{\quad \ \nu}  
\end{eqnarray} 
where $\chi_{\mu\nu}$ represents the effect of the bulk geometry
 on the brane. Importantly, this equation holds 
irrespective of the existence of the other brane. 
 Similarly,  the equation of motion on the negative tension 
brane is given by
\begin{eqnarray}
     G^{(4)\mu}_{\quad \ \nu} (f_{\mu\nu} = \Omega^2  h_{\mu\nu}) 
     =  - {\kappa^2 \over \ell}  T^{\ominus\mu}_{\quad \ \nu} 
                         + {\ell^2 \over  \Omega^4 }
                         \chi^{(1)\mu}_{\quad \ \nu} 
\end{eqnarray}
where $f_{\mu\nu}$ is the induced metric on the negative tension brane.
 Here, the effect of the bulk geometry enhanced by the factor $1/\Omega^4$
 since the bulk geometry shrinks towards to the negative tension brane. 
 Although Eqs.~(16) and (17)
are non-local individually, with undetermined $\chi^{(1)}_{\mu\nu}$,
one can combine both equations to reduce them to local equations
for each brane. This happens to be possible since
$\chi^{(1)}_{\mu\nu}$ appears only algebraically; one can easily eliminate 
$\chi^{(1)}_{\mu\nu}$ from Eqs.~(16) and (17).  
Defining a new field $\Psi = 1- \Omega^2 $,  we find 
\begin{eqnarray*}
 {\ell^3 \over 2}  \chi^{(1)\mu}_{\quad \nu} 
 &=& -{\kappa^2 (1-\Psi) \over 2 \Psi} 
      		\left( T^{\oplus \mu}_{\quad \nu} 
      		+ (1-\Psi ) T^{\ominus \mu}_{\quad \nu} \right) \\
      	      & &  -{l  \over 2 \Psi} \left[ \left(  \Psi^{|\mu}_{\ |\nu} 
  		-\delta^\mu_\nu  \Psi^{|\alpha}_{\ |\alpha} \right) 
  		+{3 \over 2(1-\Psi )} \left( \Psi^{|\mu}  \Psi_{|\nu}
  		- {1\over 2} \delta^\mu_\nu  \Psi^{|\alpha} \Psi_{|\alpha} 
  		\right) \right]     \ .
\end{eqnarray*}
The condition $\chi^{(1)\mu}_{\quad \  \mu} = 0$ gives 
 the equations of motion for the radion field:
\begin{eqnarray}
  	\Box \Psi = {\kappa^2 \over 3\ell} \left( 1-\Psi \right)
  	\left\{ T^{\oplus} + (1-\Psi) T^{\ominus} \right\}
  		-{1 \over 2 (1-\Psi ) }
  	 \Psi^{|\mu} \Psi_{|\mu} \ .
\end{eqnarray}
Interestingly, we can rearrange the above equations as
\begin{eqnarray}
 	G^\mu_{\ \nu} (h)  = {\kappa^2 \over \ell \Psi } 
 	T^{\oplus\mu}_{\quad\ \nu}
      		+{\kappa^2 (1-\Psi )^2 \over l\Psi } 
      		T^{\ominus\mu}_{\quad\ \nu}
      		+{ 1 \over \Psi } \left(  \Psi^{|\mu}_{\ |\nu} 
  		-\delta^\mu_\nu  \Psi^{|\alpha}_{\ |\alpha} \right)  
  		+{\omega(\Psi ) \over \Psi^2} \left( \Psi^{|\mu}  \Psi_{|\nu}
  		- {1\over 2} \delta^\mu_\nu  \Psi^{|\alpha} \Psi_{|\alpha} 
  		\right)  
\end{eqnarray}
and
\begin{eqnarray}
  	\Box \Psi = {\kappa^2 \over \ell} 
  	{T^{\oplus} + (1-\Psi) T^{\ominus} 
  	                                \over 2\omega (\Psi) +3}
  		-{1 \over 2\omega (\Psi) +3}
  		{d\omega (\Psi) \over d\Psi} \Psi^{|\mu} 
  		\Psi_{|\mu} \ ,
\end{eqnarray}
where the coupling function $\omega (\Psi)$ takes the following form:
$
  	\omega (\Psi ) = {3 \Psi /2(1-\Psi )}  \ .
$
We named this system as the quasi-scalar-tensor theory. 
 Eqs.(19) and (20) can be derived from 
\begin{eqnarray}
 	 S_A  = {l \over 2 \kappa^2} \int d^4 x \sqrt{-h} 
     		\left[ \Psi R(h) 
     		- {3 \over 2(1-\Psi )} 
     		\Psi^{|\alpha} \Psi_{|\alpha} \right]  
     	     + \int d^4 x \sqrt{-h} {\cal L}^{\oplus} 
      		+ \int d^4 x \sqrt{-h} \left(1-\Psi \right)^2 
      		{\cal L}^{\ominus}  \ .
\end{eqnarray}
This is  the effective action on the positive tension brane.

The effective action on the negative tension brane
 can be also derived in the similar way as
\begin{eqnarray}
 	S_B  = {l \over 2 \kappa^2} \int d^4 x \sqrt{-f} 
     		\left[ \Phi R(f) + {3 \over 2(1+ \Phi )} 
     		\Phi^{;\alpha} \Phi_{;\alpha} \right] 
    	     + \int d^4 x \sqrt{-f} {\cal L}^{\ominus}
    + \int d^4 x \sqrt{-f} (1+\Phi)^2 {\cal L}^{\oplus} \ ,
\end{eqnarray}
where $\displaystyle{\Phi = {1/ \Omega^2} -1}$.

Thus, we have derived a closed set of equations (19) and (20).
 By solving these equations, we can know the anisotropic stress
 $\chi^{(1)}_{\mu\nu}$ explicitly. 
 Now, we can make a precise predictions on the CMB fluctuations!
 This will be reported somewhere else.

\section{Conclusion}

We have developed the general formalism to obtain the effective
 action in the low energy regime. 
 
 In the case of the
single brane model, by imposing asymptotically AdS boundary condition,
 we have obtained the Einstein theory with corrections represented by
 CFT and higher curvature polynomial. It is suggested that
 the cosmological perturbation theory in the brane world can be 
 formulated as the integro-differential equations. 

 In the case of the two-brane model, we have shown that the system
 is described by the quasi-scalar-tensor theory. Equivalently, it can be
 regarded as the Einstein theory with the extra energy source, 
 $\chi^{(1)}_{\mu\nu}$ corresponding to the dark radiation 
 in the homogeneous cosmological case. This turns out to be determined
 by the radion field and the energy momentum tensors on positive and 
 negative tension branes. The next order corrections due to Kalza-Klein
 massive modes can be represented by the higher curvature terms. 
 It is interesting to study the cosmological scenario based on the
 effective action we have derived~\cite{kanno4}.

\vskip 0.5cm
\noindent
{\Large Acknowledgements}\\
This work was supported in part by the Monbukagakusho Grant-in-Aid 
for Scientific Research, Nos.~14540258.

\end{document}